\documentclass{elsarticle}
\usepackage{xcolor}
\usepackage{graphicx}
\usepackage{amsmath}   
\usepackage{amsfonts}  
\usepackage{amssymb}   
\usepackage{multirow} 
\usepackage{amssymb}
\usepackage{csquotes}
\usepackage{pifont}
\usepackage{hyperref}  

\begin{document}
\begin{frontmatter}
\title{Dual-Encoder Transformer-Based Multimodal Learning for Ischemic Stroke Lesion Segmentation Using Diffusion MRI}

\author{
Muhammad Usman$^{1}$,
Azka Rehman$^{2}$,
Muhammad Mutti Ur Rehman$^{3}$,
Abd Ur Rehman$^{4}$,
Muhammad Umar Farooq$^{5}$
}

\address{%
$^{1}$ \quad Department of Anesthesiology, Perioperative and Pain Medicine, Stanford University, CA 94305, USA \\}

\address{%
$^{2}$ \quad Department of Biomedical Sciences, Seoul National University, Seoul, 08826, South Korea \\}
\address{%
$^{3}$ \quad Department of Computer Engineering, National University of Sciences and Technology (NUST), Islamabad, Pakistan
}
\address{%
$^{4}$ \quad Department of Computer Science, The University of Alabama, Tuscaloosa, AL 35487, USA \\}

\address{%
$^{5}$ \quad Department of Computer Science, Hanyang University, Seoul, 04762, South Korea \\}


\begin{abstract}
Accurate segmentation of ischemic stroke lesions from diffusion magnetic resonance imaging (MRI) is essential for timely clinical decision-making, treatment planning, and longitudinal outcome assessment. Diffusion-Weighted Imaging (DWI) and Apparent Diffusion Coefficient (ADC) scans provide complementary information regarding acute and sub-acute ischemic changes; however, reliable automated delineation of infarct regions remains challenging due to variability in lesion size, shape, and location.

In this study, we investigate ischemic stroke lesion segmentation using multimodal diffusion MRI from the ISLES 2022 dataset. We first benchmark several state-of-the-art convolutional and transformer-based segmentation architectures, including U-Net variants, Swin-UNet, and TransUNet. Based on comparative performance, we further propose a dual-encoder TransUNet architecture designed to explicitly model modality-specific representations from DWI and ADC images. To enhance spatial contextual awareness, we incorporate adjacent slice information using a three-slice input configuration.

All models are trained using a unified preprocessing and optimization strategy and evaluated using the Dice Similarity Coefficient (DSC). Experimental results demonstrate that transformer-based architectures outperform conventional convolutional models. The proposed dual-encoder TransUNet achieves consistent performance gains over single-encoder early-fusion approaches, with the three-slice dual-encoder configuration achieving the highest segmentation accuracy, attaining a Dice score of 85.4\% on the test set.

These findings highlight the importance of effective multimodal fusion and contextual representation for ischemic stroke lesion segmentation. The proposed framework provides a robust and scalable solution for automated infarct segmentation from diffusion MRI and establishes a strong baseline for future multimodal neuroimaging research.
\end{abstract}

\begin{keyword}

Ischemic stroke \sep lesion segmentation \sep diffusion-weighted imaging \sep apparent diffusion coefficient \sep transformer-based networks

\end{keyword}
\end{frontmatter}

\section{Introduction}
\label{sec:introduction}

Ischemic stroke remains one of the leading causes of mortality and long-term disability worldwide, imposing a substantial clinical and socioeconomic burden. Early and accurate identification of infarcted tissue is essential for treatment selection, prognosis estimation, and longitudinal disease management. Diffusion-weighted imaging (DWI) and apparent diffusion coefficient (ADC) MRI play a central role in acute and sub-acute stroke assessment due to their high sensitivity to early ischemic changes. However, manual delineation of ischemic lesions from diffusion MRI is labor-intensive and subject to inter- and intra-observer variability, motivating the development of reliable automated segmentation approaches.

Over the past decade, deep learning has emerged as a powerful paradigm for medical image analysis, enabling substantial improvements across segmentation, detection, and classification tasks. Convolutional neural networks (CNNs) have demonstrated strong performance in a wide range of anatomical and pathological segmentation problems, including mandibular canal segmentation in CBCT scans, lung nodule analysis, and brain tumor segmentation \cite{usman2022dual,usman2020volumetric,rehman2023selective,ullah2022cascade}. These approaches commonly integrate attention mechanisms, deep supervision, and multi-scale feature extraction to enhance boundary localization and robustness to anatomical variability.

Beyond conventional single-encoder CNN architectures, increasing attention has been directed toward multi-encoder and dual-encoder designs, which explicitly disentangle feature learning pathways for heterogeneous inputs. Such architectures have been shown to be particularly effective in scenarios involving multimodal data or complex appearance variations. For example, dual-encoder and multi-encoder networks have achieved improved performance in lung nodule segmentation and detection through adaptive attention mechanisms, self-distillation, and multi-view fusion strategies \cite{usman2023deha,usman2024meds,usman2025multi,latif2018mobile}. These findings suggest that naive early fusion via channel concatenation may limit representational capacity when different modalities encode complementary clinical information.

Attention mechanisms and adaptive region-of-interest (ROI) modeling have further advanced the state of the art in medical image segmentation by suppressing irrelevant background information and focusing learning on clinically meaningful regions. Hard attention, soft attention, and cascaded attention strategies have been successfully applied across diverse imaging tasks, including brain tumor segmentation, lung imaging, and COVID-19 detection from chest X-rays \cite{rehman2023selective,ullah2023densely,ullah2022cascade}. Semi-supervised and multi-task learning paradigms have also demonstrated promise in improving generalization under limited annotation settings \cite{ullah2023mtss,ullah2023ssmd}.

In parallel, the adoption of transformer-based architectures has introduced new opportunities for modeling long-range spatial dependencies that are difficult to capture using purely convolutional designs. Hybrid CNN-transformer models, such as TransUNet and Swin-based architectures, have shown strong performance in medical image segmentation by combining local feature extraction with global self-attention. Related efforts have also explored lightweight and deployment-friendly models suitable for IoT and edge environments, highlighting the importance of efficiency alongside accuracy \cite{iqbal2023ldmres,usman2024intelligent,farooq2025gdssa}. Transformer-based representation learning has additionally been applied to anatomy-guided segmentation and multimodal fusion tasks in ultrasound and neuroimaging \cite{farooq2025anatomy,usmancomplex}.

Despite these advances, ischemic stroke lesion segmentation presents unique challenges. Lesions often exhibit high variability in size, shape, and spatial distribution, and diffusion MRI is susceptible to motion artifacts and intensity heterogeneity. Prior work has explored motion correction and image reconstruction in multishot MRI using generative and adversarial learning frameworks \cite{usman2020retrospective,latif2018automating}, yet systematic studies focusing on architectural design choices for multimodal diffusion MRI segmentation remain relatively limited.

More broadly, artificial intelligence has demonstrated transformative potential across biomedical signal processing, imaging, and healthcare systems. Applications span phonocardiographic sensing \cite{latif2018phonocardiographic}, affective computing \cite{usman2017using,latif2018cross}, COVID-19 analytics \cite{latif2020leveraging,ullah2023densely,ullah2023mtss}, explainable computer-aided diagnosis \cite{lee2021evaluation}, and image source–agnostic pathology pipelines \cite{usman20251575}. Multimodal representation learning has also been increasingly explored in neuroimaging for tasks such as brain age estimation and biological aging analysis \cite{usman2024advancing,rehman2024biological}. These works collectively emphasize the importance of robust, interpretable, and clinically aligned AI systems.

Motivated by these developments, this study focuses on automated ischemic stroke lesion segmentation using multimodal diffusion MRI from the ISLES 2022 dataset. We conduct a systematic benchmarking of state-of-the-art CNN- and transformer-based segmentation architectures and investigate the impact of architectural design on segmentation performance. Building upon prior successes of multi-encoder and attention-based models, we introduce a dual-encoder TransUNet framework that explicitly models modality-specific representations from DWI and ADC scans. Furthermore, we incorporate limited three-dimensional contextual information through adjacent slice aggregation to improve lesion boundary delineation.

The main contributions of this work are summarized as follows:
\begin{itemize}
    \item A comprehensive benchmarking of state-of-the-art segmentation models for ischemic stroke lesion segmentation on the ISLES 2022 dataset.
    \item A dual-encoder transformer-based architecture for effective multimodal fusion of DWI and ADC images.
    \item An empirical analysis of the role of multi-slice contextual information in improving segmentation accuracy.
\end{itemize}

Through extensive experimental evaluation, this work provides insights into effective multimodal representation learning strategies for diffusion MRI and establishes a strong foundation for future research in ischemic stroke imaging and automated clinical decision support.
\section{Methodology}
\label{sec:methodology}

\subsection{Problem Formulation}

The objective of this work is to perform automated segmentation of ischemic stroke lesions from diffusion magnetic resonance imaging (MRI). Given paired Diffusion-Weighted Imaging (DWI) and Apparent Diffusion Coefficient (ADC) scans, the task is formulated as a binary semantic segmentation problem, where each pixel is classified as either lesion or background. By leveraging complementary diffusion information from DWI and ADC modalities, the goal is to improve robustness and accuracy in delineating infarct regions across acute and sub-acute stroke cases.

\subsection{Dataset Description}

This study utilizes the ISLES 2022 dataset, introduced as part of the MICCAI 2022 Ischemic Stroke Lesion Segmentation Challenge. The dataset comprises multi-center MRI scans acquired from patients with acute and sub-acute ischemic stroke, with expert-annotated ground-truth lesion masks. Each subject includes DWI and ADC volumes, which are widely used in clinical practice for identifying early ischemic changes.

The publicly available portion of the dataset consists of 250 cases. Since the official test labels are not released, the dataset was randomly divided into 200 scans for training and 50 scans for testing. The identifiers corresponding to each split were stored in JSON files to ensure reproducibility.

\subsection{Preprocessing Pipeline}

All MRI volumes were preprocessed using a unified pipeline to ensure consistency across subjects. DWI, ADC, and segmentation masks were loaded using the \texttt{nibabel} library. Intensity normalization was applied independently to DWI and ADC volumes using min--max normalization to scale voxel intensities to the range $[0,1]$.

To remove irrelevant background regions and reduce computational overhead, spatial cropping was performed using a bounding box derived from the non-zero regions of the DWI volume. The same bounding box was applied to the corresponding ADC and mask volumes to preserve spatial alignment across modalities.

Rather than processing full 3D volumes, a slice-wise representation was adopted. To incorporate limited three-dimensional contextual information while maintaining computational efficiency, three consecutive axial slices were stacked per modality. As a result, each input sample had a spatial resolution of $128 \times 128$ with three slices per modality, yielding an input tensor of size $128 \times 128 \times 3 \times 2$. The first and last slices of each volume were excluded to ensure valid three-slice stacks. Slices with negligible signal content were discarded based on an intensity threshold. Each processed input and corresponding 2D lesion mask were stored in NumPy format for efficient training.

\subsection{Baseline Segmentation Models}

To establish benchmark performance on the ISLES 2022 dataset, several state-of-the-art segmentation architectures were evaluated, including U-Net with ResNet152 backbone, U-Net with EfficientNet backbone, Swin-UNet, and TransUNet. All baseline models were trained under identical preprocessing, data augmentation, and optimization settings to enable fair comparison. Among the evaluated architectures, TransUNet demonstrated superior performance and was therefore selected as the base architecture for further customization.

\subsection{Customization of TransUNet Architectures}
\label{subsec:transunet_customization}

\begin{figure}[t]
    \centering
    \includegraphics[width=\linewidth]{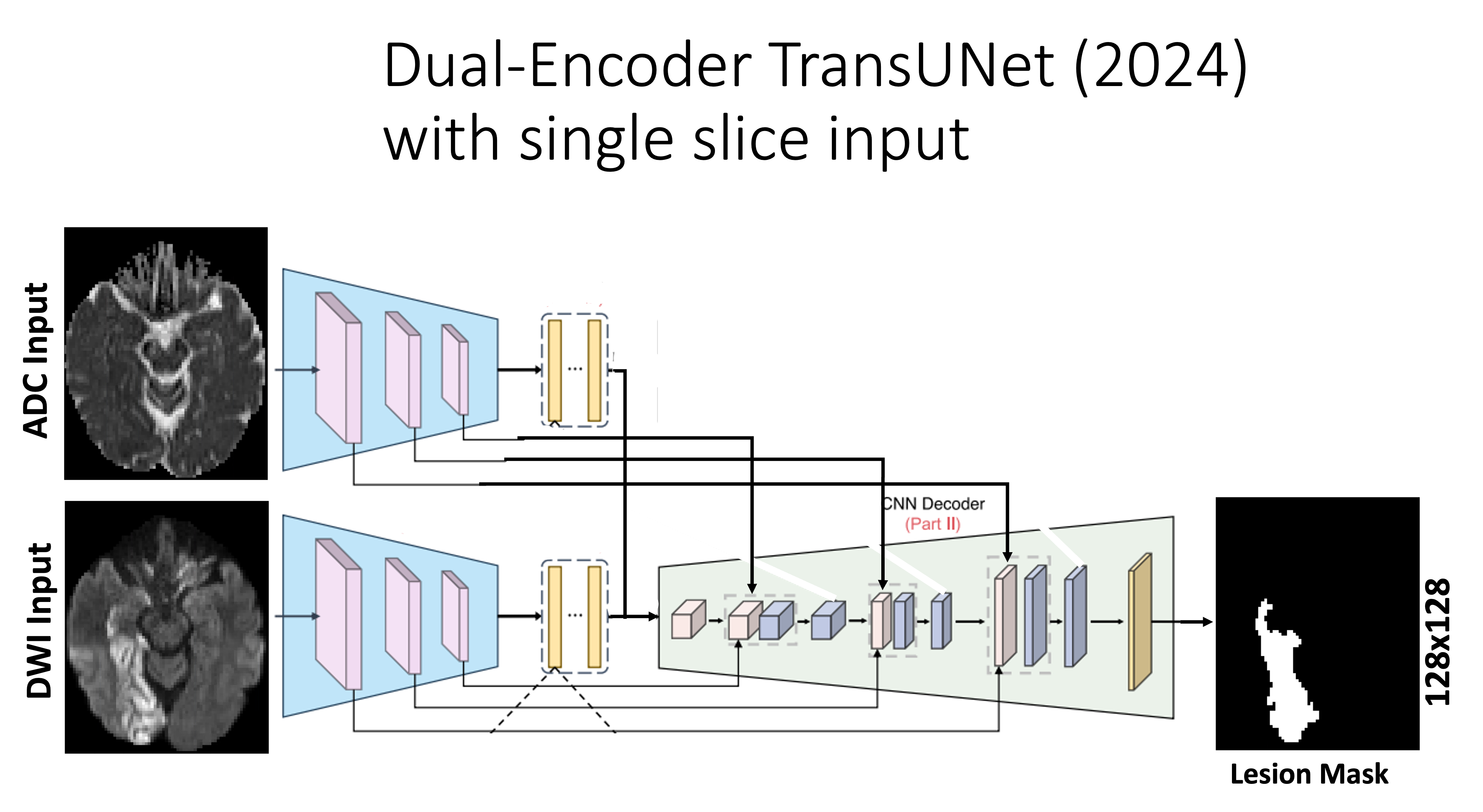}
    \caption{Dual-encoder TransUNet with single-slice input. DWI and ADC are processed by independent encoders, and modality-specific features are fused at the bottleneck before transformer decoding.}
    \label{fig:transunet_dual_single}
\end{figure}

\begin{figure}[t]
    \centering
    \includegraphics[width=\linewidth]{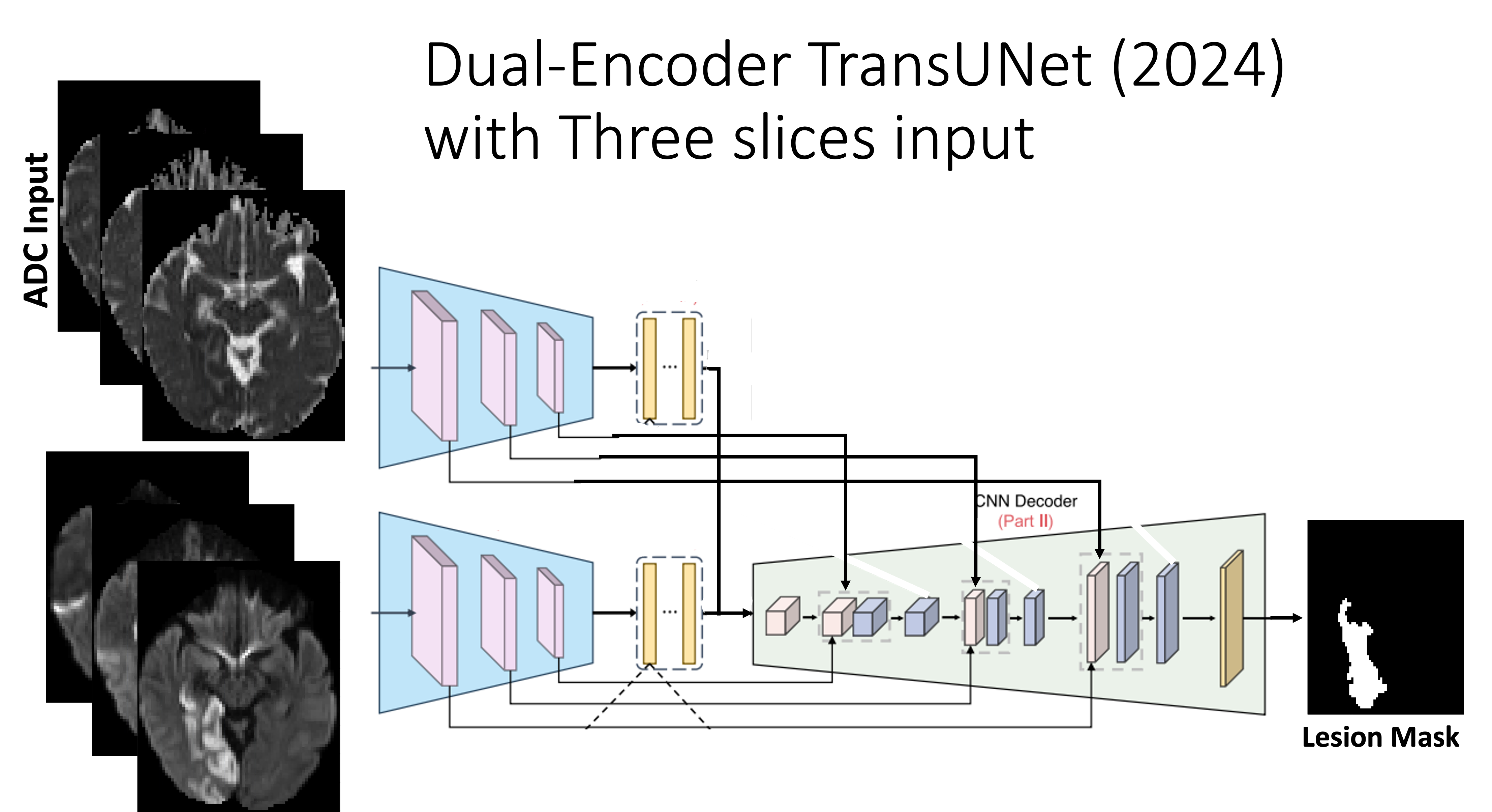}
    \caption{Dual-encoder TransUNet with three-slice input. Three consecutive slices per modality provide limited inter-slice context, improving lesion boundary delineation while maintaining slice-wise training efficiency.}
    \label{fig:transunet_dual_three}
\end{figure}

To better exploit multimodal diffusion MRI information and spatial context, three TransUNet-based configurations were investigated, as illustrated in Fig.~\ref{fig:transunet_single}--Fig.~\ref{fig:transunet_dual_three}.

\subsubsection{Single-Encoder TransUNet}
\begin{figure}[t]
    \centering
    \includegraphics[width=\linewidth]{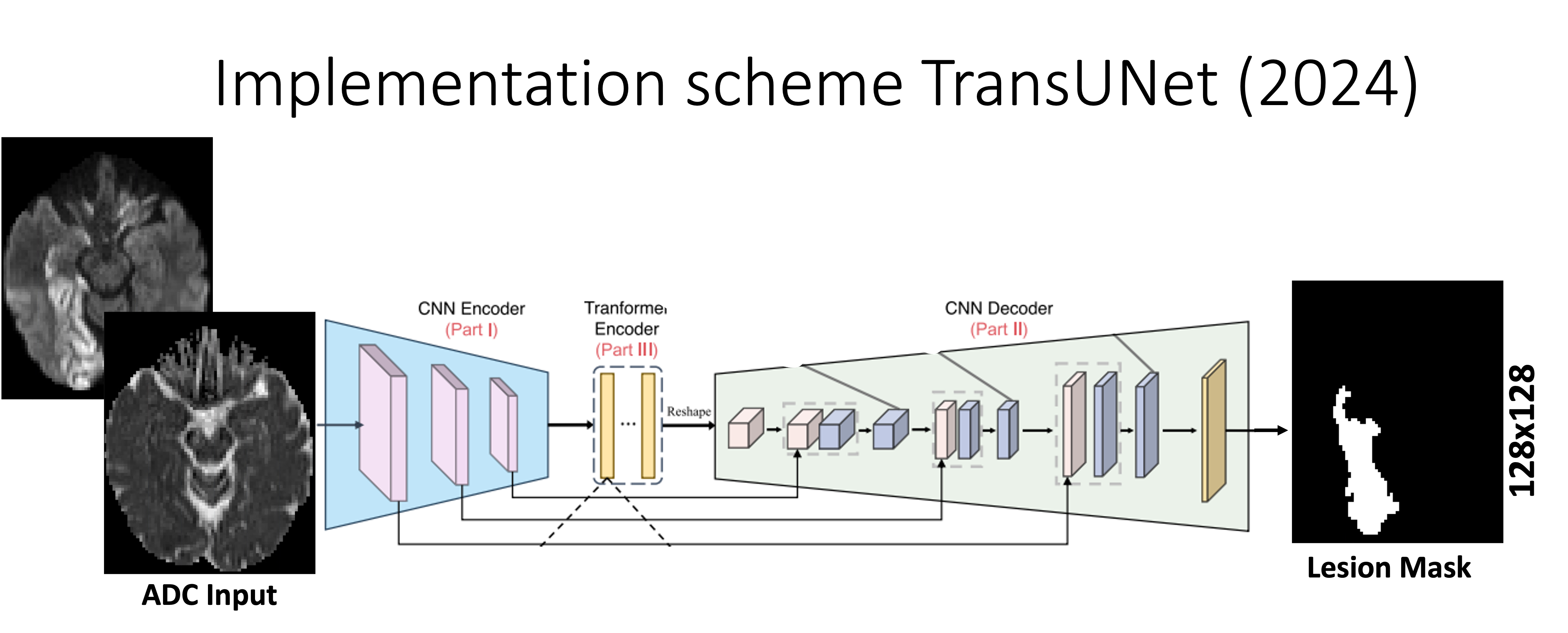}
    \caption{Single-encoder TransUNet architecture for multimodal ischemic stroke lesion segmentation. DWI and ADC inputs are fused via early concatenation and processed through a shared encoder, transformer module, and decoder.}
    \label{fig:transunet_single}
\end{figure}

The first configuration corresponds to the original single-encoder TransUNet architecture, where DWI and ADC inputs are fused through early channel-wise concatenation and processed by a shared convolutional encoder, followed by a transformer module and a decoder network (Fig.~\ref{fig:transunet_single}). This design enables joint feature learning across modalities but may limit the model’s ability to capture modality-specific characteristics.

\subsubsection{Dual-Encoder TransUNet with Single-Slice Input}

To explicitly model modality-specific representations, a dual-encoder TransUNet architecture was introduced. In this configuration, DWI and ADC images are processed through two independent encoders with identical architectures but separate parameters. The learned feature maps are fused at the bottleneck stage before being passed to the transformer and decoder components (Fig.~\ref{fig:transunet_dual_single}). This late-fusion strategy allows the network to preserve complementary information from each modality while improving feature discrimination.

\subsubsection{Dual-Encoder TransUNet with Three-Slice Input}

Finally, the dual-encoder architecture was extended to incorporate three consecutive axial slices per modality (Fig.~\ref{fig:transunet_dual_three}). This configuration provides limited inter-slice contextual information, which is particularly beneficial for accurately delineating lesion boundaries and handling small or fragmented infarcts. Despite leveraging additional spatial context, this approach retains a slice-wise training paradigm and avoids the computational burden of full 3D processing.

\subsection{Training Strategy}

All models were trained using a unified optimization strategy. Binary Cross-Entropy with Logits loss was employed to supervise voxel-wise predictions. Training was performed with a batch size of 16 for a total of 100 epochs. A two-stage training scheme was adopted, consisting of an initial phase with frozen encoder layers for five epochs, followed by full network fine-tuning. The best-performing model was selected based on validation loss.

\subsection{Data Augmentation}

To improve generalization, extensive data augmentation was applied during training, including random horizontal and vertical flipping, random rotations up to 270 degrees, and resizing all inputs to a fixed resolution of $128 \times 128$. Augmentations were applied consistently to both images and corresponding segmentation masks.

\subsection{Implementation Details}

All experiments were implemented using PyTorch, with FastAI employed for training U-Net–based models. The TransUNet architectures were implemented using the official codebase and extended to support multimodal and multi-slice inputs. Training progress, loss curves, and evaluation metrics were monitored using Weights \& Biases to ensure reproducibility and systematic experiment tracking.
\section{Results}
\label{sec:results}

This section presents a quantitative and qualitative evaluation of the proposed ischemic stroke lesion segmentation framework. We first establish baseline performance using several state-of-the-art segmentation architectures on the ISLES 2022 dataset. We then analyze the effects of multimodal fusion strategies, encoder design, and contextual slice aggregation on segmentation performance.

\subsection{Evaluation Metrics}

Segmentation performance was primarily evaluated using the Dice Similarity Coefficient (DSC), a widely adopted metric for assessing spatial overlap between predicted segmentation masks and ground-truth annotations. The DSC is defined as:
\[
\text{DSC} = \frac{2 |P \cap G|}{|P| + |G|}
\]
where $P$ denotes the set of predicted lesion pixels and $G$ denotes the set of ground-truth lesion pixels.

The DSC ranges from 0 to 1, with higher values indicating better agreement between prediction and ground truth. This metric is particularly appropriate for ischemic stroke lesion segmentation, where both false positives and false negatives have important clinical implications.

\subsection{Baseline Model Comparison}

To establish benchmark performance on the ISLES 2022 dataset, multiple state-of-the-art segmentation models were evaluated under identical training, preprocessing, and augmentation settings. The evaluated architectures included U-Net variants with ResNet152 and EfficientNet backbones, Swin-UNet, and TransUNet.

Quantitative results are summarized in Table~\ref{tab:baseline_results}. Among all baseline models, TransUNet achieved the highest Dice score, indicating superior segmentation accuracy compared to both convolutional and transformer-based alternatives.

\begin{table}[ht]
\centering
\caption{Baseline segmentation performance on the ISLES 2022 test set.}
\label{tab:baseline_results}
\begin{tabular}{lc}
\hline
Model & Dice Score (\%) \\
\hline
U-Net (ResNet152) & 74.2 \\
U-Net (EfficientNet) & 71.6 \\
Swin-UNet & 76.4 \\
TransUNet & 81.3 \\
\hline
\end{tabular}
\end{table}

The improved performance of TransUNet suggests that its hybrid CNN-transformer architecture effectively captures both local texture information and long-range contextual dependencies in diffusion MRI.

\subsection{Effect of Multimodal Fusion and Encoder Design}

To investigate the impact of multimodal feature fusion, several TransUNet variants were evaluated using combined DWI and ADC inputs. The original single-encoder TransUNet architecture, which performs early fusion through channel concatenation, achieved a Dice score of 81.3\%.

To further enhance modality-specific feature learning, a dual-encoder TransUNet architecture was introduced. In this configuration, DWI and ADC modalities were processed through independent encoders, followed by feature fusion at the bottleneck stage. Using single-slice inputs, the dual-encoder TransUNet achieved a Dice score of 83.1\%, representing a noticeable improvement over the single-encoder baseline.

These results indicate that late fusion of multimodal representations is more effective than early fusion, particularly when imaging modalities capture distinct pathological characteristics of ischemic lesions.

\subsection{Impact of Multi-Slice Contextual Information}

The influence of spatial context was further examined by extending the dual-encoder TransUNet to incorporate three consecutive axial slices per modality. This configuration enables the network to leverage limited three-dimensional contextual information while preserving a slice-wise training framework.

The three-slice dual-encoder TransUNet achieved the highest segmentation performance, with a Dice score of 85.4\%. Compared to the single-slice dual-encoder variant, this represents a consistent and meaningful improvement, underscoring the importance of inter-slice continuity for accurate lesion boundary delineation.

A summary of TransUNet-based experimental results is provided in Table~\ref{tab:custom_results}.

\begin{table}[ht]
\centering
\caption{Performance comparison of TransUNet variants using DWI and ADC inputs.}
\label{tab:custom_results}
\begin{tabular}{lc}
\hline
Model Configuration & Dice Score (\%) \\
\hline
TransUNet (Single Encoder, Single Slice) & 81.3 \\
Dual-Encoder TransUNet (Single Slice) & 83.1 \\
Dual-Encoder TransUNet (Three Slices) & 85.4 \\
\hline
\end{tabular}
\end{table}

\subsection{Discussion of Experimental Findings}

The experimental results highlight three key observations. First, hybrid CNN-transformer architectures outperform purely convolutional models for ischemic stroke lesion segmentation, likely due to their ability to model global spatial relationships. Second, explicitly separating DWI and ADC feature extraction through a dual-encoder design improves multimodal representation learning and segmentation accuracy. Finally, incorporating adjacent slice context substantially enhances performance, particularly for small, fragmented, or irregularly shaped infarcts that are difficult to delineate using single-slice information.

Overall, these findings demonstrate that careful architectural design and multimodal contextual modeling are critical for achieving robust and accurate ischemic stroke lesion segmentation from diffusion MRI.
\section{Conclusion}
\label{sec:conclusion}

In this study, we investigated ischemic stroke lesion segmentation from diffusion MRI by leveraging multimodal information from Diffusion-Weighted Imaging (DWI) and Apparent Diffusion Coefficient (ADC) scans. Using the ISLES 2022 dataset, we conducted a systematic evaluation of state-of-the-art segmentation architectures and explored architectural modifications to improve robustness and accuracy in acute and sub-acute stroke settings.

Through extensive benchmarking, transformer-based hybrid models were shown to outperform conventional convolutional architectures, with TransUNet achieving the strongest baseline performance. Building upon this observation, we introduced a dual-encoder TransUNet architecture to enable modality-specific feature extraction from DWI and ADC inputs. This design consistently improved segmentation accuracy over single-encoder early-fusion approaches, highlighting the importance of preserving complementary information across imaging modalities.

Furthermore, incorporating limited three-dimensional contextual information via multi-slice input significantly enhanced segmentation performance. The proposed dual-encoder TransUNet with three-slice DWI and ADC inputs achieved the highest Dice score, demonstrating improved delineation of lesion boundaries, particularly for small and spatially fragmented infarcts.

Overall, the results indicate that both multimodal fusion strategy and contextual representation play critical roles in ischemic stroke lesion segmentation. The proposed framework provides a robust and effective solution for automated infarct segmentation from diffusion MRI and establishes a strong baseline for future research on multimodal stroke imaging.

Future work will focus on extending the proposed approach to full volumetric segmentation, evaluating model generalizability across external cohorts, and incorporating additional imaging modalities such as FLAIR or perfusion MRI. Integrating uncertainty estimation and clinically relevant outcome prediction may further enhance the translational potential of the proposed framework in real-world stroke assessment and decision support systems.

\bibliographystyle{elsarticle-num}
\bibliography{refs.bib}

\end{document}